  \documentclass[final,3p,times,twocolumn]{elsarticle}
\usepackage{amssymb}
\usepackage{amsmath,amssymb,exscale}
\usepackage{graphicx}
\usepackage{epsfig}
\usepackage{multicol}
\usepackage{color}
\usepackage{mathrsfs}
\usepackage{hyperref}
\hypersetup{colorlinks,bookmarksopen,bookmarksnumbered,citecolor=rossos,
linkcolor=black,pdfstartview=FitH,urlcolor=blus}
\usepackage{amsfonts}
\usepackage{slashed}
\usepackage{blindtext}
 \usepackage{fancyhdr}
\usepackage{hyperref}

\definecolor{blus}{cmyk}{1,1,0,0.6}
\definecolor{verdes}{cmyk}{0.99,0,0.99,0.02}
\definecolor{rossos}{cmyk}{0,1,1,0.55}
\definecolor{greeny}{cmyk}{0.99,0,0.59,0.98}

\def\bp{\bar M_{\rm Pl}}
\def\Lag{\mathscr{L}}

\def\be{\begin{equation}}
\def\ee{\end{equation}}
\def\bea{\begin{eqnarray}}
\def\eea{\end{eqnarray}}
\def\ba{\begin{array} }
\def\ea{\end{array}}
\def\bac{\begin{array} {c}}
\def\bacc{\begin{array} {cc}}
\def\baccc{\begin{array} {ccc}}

\definecolor{red}{rgb}{1,0,0}

\newcommand{\mub}{\bar{\mu}}

\def\hhref#1{\href{http://arxiv.org/abs/#1}{arXiv:#1}} % in bibliography

\journal{the arXiv}

\begin{document}

\begin{frontmatter}



%Preprint UAM:  FTUAM-13-22
%Preprint IFT: IFT-UAM/CSIC-13-089

\title{\vspace{-2cm}\huge Initial Conditions for Critical  Higgs Inflation}

%\author{Alberto Salvio}

\author{\vspace{1cm}{\large Alberto Salvio}}

\address{\normalsize \vspace{0.2cm}CERN, Theoretical Physics Department, Geneva, Switzerland.  

\vspace{0.3cm}
 {\it {\small Report number: CERN-TH-2017-261}}
 \vspace{-1cm}
 }

\begin{abstract}
\noindent   It has been pointed out that a large non-minimal coupling $\xi$ between the Higgs and the Ricci scalar can source higher derivative operators, which may change the predictions of Higgs inflation. A variant, called critical Higgs inflation, employs the near-criticality of the top mass to introduce an inflection point in the potential and lower drastically the value of $\xi$. We here study whether critical Higgs inflation can occur even if the pre-inflationary initial conditions do not satisfy the slow-roll behaviour (retaining translation and rotation symmetries). A positive answer is found: inflation turns out to be an attractor and therefore no fine-tuning of the initial conditions is necessary. A very large initial Higgs time-derivative (as compared to the potential energy density) is compensated by a moderate increase in the initial field value. These conclusions are reached by solving the exact Higgs equation without using the slow-roll approximation. This also allows us to consistently treat the inflection point, where  the standard slow-roll approximation breaks down. Here we make use of an approach that is independent of the UV completion of gravity, by taking initial conditions that always involve sub-planckian energies. 
\end{abstract}


\end{frontmatter}

\section{Introduction}\label{introduction}

\vspace{-0.1cm}

\noindent So far no clear evidence for physics beyond the Standard Model at the scales explored by the LHC has been found. In this situation, it is useful to look for complementary tests. The extrapolation of the SM at energies much above those reachable at colliders offers a new way to look for further evidence of new physics (besides the already established ones, such as neutrino oscillations and dark matter). 

 Inflation is a natural arena to perform these tests. It was found that the Higgs of the SM might play the role of the inflaton provided that a sizable non-minimal coupling $\xi$ with the Ricci scalar $R$ is introduced. Ref.~\cite{Bezrukov:2007ep} considered originally the case of a very large $\xi$, which corresponds to the SM living well inside the so called stability region\footnote{This is  the region of parameter space where the electroweak (EW) vacuum is the global minimum of the SM effective potential.}. In this setup two different scales appear, the reduced Planck mass $\bp\simeq 2.435 \times 10^{18}$~GeV and $\bp/\xi$ and a violation of perturbative unitarity at  $\bp/\xi$ has been found  by considering scatterings of particles viewed as fluctuations around the EW vacuum~\cite{crit,Burgess:2010zq}. This leads to the necessity of new physics or strong coupling methods to analyse that physical situation. While this does not undoubtedly exclude Higgs inflation (HI) as the relevant expansion in that case is around a large Higgs field~\cite{Bezrukov:2010jz}, the issue may be avoided by living very close to the boundary between stability and metastability\footnote{The metastability region is the region of parameter space where the lifetime of the EW vacuum exceeds the age of the universe.}, the so called criticality. Indeed, a drastic decrease of $\xi$ occurs at criticality~\cite{Hamada:2014iga,Bezrukov:2014bra,Hamada:2014wna}, leading to a single new physics scale, $\bp$, where quantum gravity effects are expected to emerge.

 Moreover, in~\cite{Salvio:2015kka}  another issue of the large-$\xi$ HI was pointed out. At the quantum level, it is necessary to tune  the high energy values of some parameters in order to preserve the inflationary predictions: if this is not done large higher derivative terms in the effective action, such as $R^2$, are generated, changing the output of the model (see also~\cite{Kannike:2015apa}). However, Ref.~\cite{Salvio:2015kka} did not consider the critical HI case, which, as stated above, does not require a large $\xi$.
 %READ UP TO HERE
 
 The aim of this article is to investigate whether critical HI suffers from any tuning in the choice of the high energy parameters. This will include in particular  the analysis of the dependence of critical HI on the initial (pre-inflationary) conditions. Indeed, {\it any} slow-roll model of inflation, such as HI, should provide a mechanism that drives generic initial conditions to slow-rolling configurations, i.e. an inflationary attractor. If such an attractor does not exist a fine-tuning of the initial conditions is required, which makes the whole idea of inflation less attractive, given that its main purpose is to solve  fine-tuning problems (the horizon and flatness problems).  
 
The paper is organized as follows. In Sec.~\ref{model} details of HI are given, including the classical analysis of inflation and a description of the quantum corrections; there, we will address the question of whether a fine-tuning of the high energy conditions of the running parameters is required in critical HI.  In Sec.~\ref{Pre-inflationary dynamics} we will consider initial conditions violating the slow-roll behaviour in order to establish the existence of an inflationary attractor in HI; both analytical and numerical arguments will be used. Finally, Sec.~\ref{Conclusions} provides the conclusions.

\vspace{-0.2cm}

%%CHECKED SPELLING OF TITLE, ABSTRACT AND CONCLUSION AND UP TO HERE 
\section{The model}\label{model}
 
 \vspace{-0.1cm}
Let us define the Higgs inflation model \cite{Bezrukov:2007ep}. The action is 
\begin{equation} S= \int d^4x\sqrt{-g}\left[\Lag_{\rm SM}-\left(\frac{\bp^2 }{2}+\xi |H|^2\right)R\right], \label{Jordan-frame-total}\end{equation}
where $H$ is the Higgs doublet, $\xi$ is a real parameter and $\sqrt{-g}\Lag_{\rm SM}$ is the SM Lagrangian minimally coupled to gravity. 
The part of the action that depends on the metric and the Higgs field  {\it only} (the scalar-tensor part) is 
\begin{equation} S_{\rm st} = \int d^4x\sqrt{-g}\left[|\partial H|^2-V-\left(\frac{\bp^2 }{2}+\xi |H|^2\right)R\right], \label{Jordan-frame}\end{equation}
where  
 $V=\lambda (|H|^2-v^2/2)^2$ is the classical Higgs potential, and $v$ is the EW Higgs vacuum expectation value.  We assume a sizable non-minimal coupling, $\xi>1$, because this is required by inflation as we will see.

\subsection{Classical analysis}\label{Classical analysis}

The $\xi |H|^2 R$ term can be eliminated through a {\it conformal} transformation (a.k.a. Weyl transformation):
\begin{equation} g_{\mu \nu}\rightarrow   \Omega^{-2}  g_{\mu \nu}, \quad \Omega^2= 1+\frac{2\xi |H|^2}{\bp^2}. \label{transformation}\end{equation}
The original frame, where the Lagrangian has the form in Eq.~(\ref{Jordan-frame-total}), is called the Jordan frame, while the one where gravity is canonically normalized (obtained with the transformation above) is called the Einstein frame.
In the unitary gauge, where the only scalar field is the radial mode $\phi \equiv \sqrt{2|H|^2}$,  we have (after  having performed the conformal transformation)
\begin{equation} S_{\rm st} = \int d^4x\sqrt{-g}\left[K \frac{(\partial \phi)^2}{2}-\frac{V}{\Omega^4}-\frac{\bp^2 }{2}R\right], \label{Sst}\end{equation}
and 
\be K =  \Omega^{-4} \left[\Omega^2 +\frac32 \left(\frac{d\Omega^2}{d\phi}\right)^2 \right]. \label{K1} \ee

The non-canonical Higgs kinetic term can be made canonical through the  field redefinition $\phi=\phi(\chi)$ defined by
\begin{equation} \frac{d\chi}{d\phi}= \Omega^{-2} \sqrt{\Omega^2 +\frac32 \left(\frac{d\Omega^2}{d\phi}\right)^2 } ,\label{chi}\end{equation}
with the conventional condition $\phi(\chi=0)=0$. Note that $\phi(\chi)$ is invertible because Eq.~(\ref{chi}) tells us $d\chi/d\phi> 0$. 
%One can find a closed expression of $\chi$ as a function of $\phi$:
%\bea \chi(\phi) =  \bp \sqrt{\frac{1+6 \xi }{\xi}} \text{sinh}^{-1}\left[\frac{\sqrt{\xi  (1+6 \xi )}\phi }{\bp}\right] \nonumber\\ -\sqrt{6} \bp \text{tanh}^{-1}\left[\frac{\sqrt{6} \xi \phi }{\sqrt{\bp^2+\xi  (1+6 \xi )\phi ^2}}\right].\eea
Thus, one can extract the function $\phi(\chi)$ by inverting the function $\chi(\phi)$ defined above. 

Note that $\chi$ feels a potential 
\begin{equation} U\equiv \frac{V}{\Omega^4}=\frac{\lambda(\phi(\chi)^2-v^2)^2}{4(1+\xi\phi(\chi)^2/\bp^2)^2}\label{U} . \end{equation}

Let us now recall how slow-roll inflation emerges in this context. From (\ref{chi}) and (\ref{U}) it follows \cite{Bezrukov:2007ep} that $U$ is exponentially flat when $\chi \gg \bp$, which is a key property to have inflation. Indeed, for such high field values the quantities
\be \epsilon_U \equiv\frac{\bp^2}{2} \left(\frac{1}{U}\frac{dU}{d\chi}\right)^2, \quad \eta_U \equiv \frac{\bp^2}{U} \frac{d^2U}{d\chi^2} \label{epsilon-def}\ee
are guaranteed to be small. Therefore, the region in field configurations where $\chi \gtrsim \bp$ (or equivalently \cite{Bezrukov:2007ep} $\phi \gtrsim \bp/\sqrt{\xi}$) corresponds to inflation. 
In Sec.~\ref{Pre-inflationary dynamics} we will investigate whether successful sow-roll inflation emerges also for large initial field kinetic energy. 
In this subsection we simply assume that the time derivatives are small. 
In this case, during the whole inflation the slow-roll parameters $\epsilon_U$ and $\eta_U$ are small and the slow-roll approximation can be used.

All the parameters of the model can be determined with good accuracy through experiments and observations, including $\xi$ \cite{Bezrukov:2007ep, Bezrukov:2008ut}. $\xi$ can be fixed by requiring that the measured curvature power spectrum (at horizon exit\footnote{We  use a standard notation: $a$ is the cosmological scale factor, $H\equiv \dot a/a$ and a dot denotes the derivative with respect to (cosmic) time, $t$.} $q=aH$) \cite{Ade:2015xua}\footnote{See for instance Table 3 of the second paper in Ref.~\cite{Ade:2015xua} ($P_R$ is denoted with $A_s$ in that table). The value quoted here corresponds to the one with the smallest uncertainty in that table.},
\begin{equation}P_{R}(q)\simeq (2.14 \pm 0.06) \times 10^{-9} , \label{normalization} \end{equation}
is reproduced for a field value $\phi=\phi_{\rm b}$ corresponding to an appropriate number of e-folds \cite{Bezrukov:2008ut}:
\begin{equation}N=\int_{\phi_e}^{\phi_{\rm b}}\frac{U}{\bp^2}\left(\frac{dU}{d\phi}\right)^{-1}\left(\frac{d\chi}{d\phi}\right)^2d\phi\simeq 59, \label{e-folds}\end{equation}
where $\phi_e$ is the field value at the end of inflation, computed by requiring 
\be \epsilon(\phi_e) \simeq 1. \label{inflation-end}\ee
In the slow-roll approximation (used in this subsection) such constraint can be imposed by using the standard formula
\begin{equation}P_{R}(k)= \frac{U/ \epsilon_U}{24\pi^2 \bp^4}.\label{PRsr} \end{equation}
 For $N=59$, this procedure leads to 
  \be \xi \simeq (5.02\mp0.06)\times10^4 \sqrt{\lambda},\qquad (N=59)\label{xi large}\ee 
where the uncertainty corresponds to the experimental uncertainty quoted in Eq.~(\ref{normalization}). Note that $\xi$ depends on $N$:
 \bea \xi &\simeq& (4.61\mp0.06)\times10^4 \sqrt{\lambda}\qquad (N=54),\label{xi large2} \\ \xi &\simeq& (5.43\mp0.06)\times10^4 \sqrt{\lambda}\qquad (N=64).\label{xi large3}\eea 
 Given that  $\lambda \sim 0.1$,  $\xi$ has to be much larger than one at the classical level. The need of a very large $\xi$ can be avoided when quantum corrections are included \cite{Hamada:2014iga,Bezrukov:2014bra,Hamada:2014wna}, as we will see in the next subsection.
 
 We can also extract the scalar spectral index $n_s$ and  the tensor-to-scalar ratio $r$: in the slow-roll approximation we are using in this subsection the  formul$\ae$  are  $r =16\epsilon_U$ and 
$n_s=1-6\epsilon_U +2\eta_U$.
These parameters are of interest as they are constrained by observations.

 \vspace{-0.2cm}

\subsection{Quantum corrections}\label{Quantum corrections}

Here we take into account quantum corrections to the Higgs potential. We would like to include both the large-$\xi$ inflationary scenario of \cite{Bezrukov:2007ep} and the critical Higgs inflation proposed in \cite{Hamada:2014iga,Bezrukov:2014bra,Hamada:2014wna}, which employs a value of the top mass close to the frontier between stability and metastability of the electroweak vacuum. The latter case allows for a drastic decrease of the required value of $\xi$ with respect to the result of the classical analysis (see Eqs.~(\ref{xi large})-(\ref{xi large3}).). This indicates that we cannot rely on large-$\xi$ approximations to analyse this case. We therefore do not use such approximations here. However, we do assume in the following that $\xi > 1$ as this condition is present both in the original formulation of HI  and in critical HI. 

Note that Eqs.~(\ref{transformation})-(\ref{chi}) hold also if $\xi$ is field-dependent, as dictated by quantum corrections~\cite{Ezquiaga:2017fvi}.
%Note however that one should take the expression in (\ref{chi}) or the one in (\ref{chi2})  according to the definition of the theory: one may defining the quantum theory by quantizing in the Einstein frame and in that case it would make sense to use (\ref{chi}).
A second step we should do now is the computation of the  effective potential. In defining the quantum theory there are well-known ambiguities \cite{Bezrukov:2009db,Bezrukov:2009-2, Bezrukov:2014bra,Bezrukov:2014ipa,Bezrukov:2017dyv}. We follow here Ref.~\cite{Bezrukov:2014bra} and choose to determine the loop corrections to the effective potential - a.k.a. Coleman-Weinberg potential - in the Einstein frame (after having performed the conformal transformation  (\ref{transformation})); the effective potential is also RG-improved by using the RGEs\footnote{We use  dimensional regularization (DR) to regularize the loop integrals and the modified minimal subtraction ($\overline{\rm MS}$) scheme to renormalize away the divergences. This, as usual, leads to a renormalization scale, $\mub$.} of the SM properly modified to take into account $\xi$. The way $\xi$ affects the running is through the appearance of a factor $s$ that suppresses the contribution of the physical Higgs field to the RGEs \cite{DeSimone:2008ei}. Ref.~\cite{DeSimone:2008ei} found 
\be s = \frac{1+\xi \phi^2/\bp^2}{1+(1+6\xi)\xi \phi^2/\bp^2}. \label{s-form} \ee
For small enough $\phi$ one has $s \approx 1$, while in the large-$\phi$ limit $s\approx 1/(1+6\xi)$.
This result does not really depend on the size of $\xi$, but, of course, the larger $\xi$ is the more effective the  suppression is.

 Such procedure to compute the effective potential is known as Prescription I and it leads to the following value of the renormalization group scale 
\be \bar\mu(\phi) =  \frac{\phi/\kappa}{\sqrt{1+\xi \phi^2/\bp^2}}, \label{mubar}\ee
where $\kappa$ is an order one  factor. This formula follows from the fact that the loop corrections to the effective potential are determined in the Einstein frame\footnote{The explicit detailed expression of the 1-loop correction can be found in Ref.~\cite{Salvio-inf}.}. The function $\bar\mu(\phi)$ can also be inverted to obtain $\phi(\bar\mu)$ and used in (\ref{s-form}) to express $s$ as a function of $\bar\mu$ only, as appropriate for the RGEs. 

The SM RGEs  modified by the $s$-factor can be found in the appendix of Ref.~\cite{Allison:2013uaa}, where the RGE of $\xi$ is also provided. We will employ these formul$\ae$ in the numerical calculation of Sec.~\ref{Numerical studies}.

Furthermore,  we will use the RG-improved potential neglecting the loop corrections: this means that we will take as effective potential the one in (\ref{U}) with the constants $\lambda$ and $\xi$ substituted with the corresponding running parameters. There are good reasons to use this approximation. 
%\begin{itemize}
First, our main objective is to see if the initial conditions with large time-derivatives of the Higgs field are attracted towards a slow-roll regime. In order to know if there is  this {\it qualitative} behaviour we do not need a very precise determination of the effective potential (which is beyond the scope of the present work).  Note, moreover, that taking into account the loop corrections to the potential would only be more precise if supplemented by the loop corrections to the kinetic term of the inflaton; such corrections have not been included in the analysis of HI and are expected to be comparable to the loop corrections to the potential for moderate values of $\xi$, unlike what happens for large $\xi$ \cite{Bezrukov:2009-2}: the large value of $\xi$ allowed~\cite{Bezrukov:2009-2} to show that the corrections to the kinetic term in the effective action are negligible, but the smaller value of $\xi$ of critical HI does not permit to trust this approximation anymore. 
Another reason to employ the RG-improved potential is its gauge independence, which is not shared by the loop corrections to the effective potential. Therefore, the use of the RG-improved potential allows for a more transparent physical interpretation.
%\end{itemize}

As boundary conditions to solve the RGEs of the SM couplings we use the currently most precise determinations of their values at the top pole mass $M_t$, which were computed in Ref.~\cite{Buttazzo:2013uya}. These values are functions of some observables: $M_t$, the Higgs and $W$ pole masses $M_h$ and $M_W$, respectively, and  $\alpha_3(M_Z)$.
%, $g_{1,2,3}(M_t)$ and $y_t(M_t).
For $M_W$ and  $\alpha_3(M_Z)$  we take the same values quoted in Ref.~\cite{Buttazzo:2013uya}, while for  $M_h$ we take the more precise determination presented in Ref.~\cite{Aad:2015zhl}, that is $M_h = 125.09 \pm 0.21\pm 0.11$~GeV.  The boundary condition for $\xi$ is instead fixed to reproduce the experimental values of the inflationary observables. The top pole mass is a variable in this work.

Now, Eqs.~(\ref{epsilon-def}), (\ref{e-folds}) and (\ref{PRsr}) of Sec.~\ref{Classical analysis} are still valid as long as one is in the slow-roll regime, but one should now interpret $U$ as the effective potential, not just as the classical  potential. However, as we will see in Sec.~\ref{Pre-inflationary dynamics},  in critical Higgs inflation, because of the presence of an inflection point in the potential (see Fig.~\ref{running-potential2}), the standard slow-roll condition may not be always satisfied; in particular it can break down around the inflection point, where the inertial term in the inflaton equation may not be negligible with respect to the friction term \cite{Garcia-Bellido:2017mdw,Ezquiaga:2017fvi,Kannike:2017bxn,Germani:2017bcs}. We will discuss further this point in Sec.~\ref{Pre-inflationary dynamics}. Nevertheless, already at this level we can observe that the slow-roll condition is violated also  at the beginning of the inflaton path if we start from initial conditions with large time derivative of the Higgs field. Therefore, the framework to analyse the pre-inflationary dynamics with such initial condition (which will be discussed in Sec.~\ref{Pre-inflationary dynamics}) will be applicable to the period when the Higgs crossed the inflection point
%\footnote{It is interesting to note that the inflection point of critical HI leads to an amplification of curvature power spectrum after inflation with the possible production of primordial black holes, which have been proposed as dark matter candidates~\cite{Ezquiaga:2017fvi}.}
 too.
  
Having determined the effective potential we can now estimate the relevant inflationary scales. In HI the energy density during inflation $U_I$ is roughly given by\footnote{One thing we learn then is that, once the observed $P_{R}$ is reproduced,  a relation between $\lambda$ and $\xi$ emerges; it is this relation that reduces the value of $\xi$ in critical Higgs inflation: what matters in reproducing $P_R$ is only the overall constant, $\lambda/\xi^2$, therefore, a smaller $\lambda$ allows us to have a smaller $\xi$.}  $U_I\sim \lambda \bp^4/\xi^2$, as clear from Eq.~(\ref{U}) and the discussion below that formula. This is related to the inflationary Hubble scale $H_I$ through the Einstein equations, $H_I\sim \sqrt{U_I}/\bp$. The scale $H_I$ is much lower than the new physics scale $\bp/\xi$ in critical HI thanks to the  smallness of $\lambda$ at the inflationary scale.  Furthermore, the smallness of $\lambda$ also leads to a small $U_I$ in Planck units, which allows us to treat gravity classically, as we will discuss in Sec.~\ref{Pre-inflationary dynamics}.  For example, in Fig.~\ref{running-potential2}, $U_I\sim 10^{-9}\bp^4$, which gives $H_I\sim 10^{-5} \bp$, a Hubble scale much smaller than the new physics scale $\bp/\xi$ for the corresponding value of the non-minimal coupling: $\bp/\xi \sim 10^{-1}\bp$. Therefore, although new degrees of freedom or strong coupling should eventually appear at $\bp/\xi$, the relevant inflationary scale $H_I$ is lower than $\bp/\xi$.
 
We now recall an important result of  Ref.~\cite{Salvio:2015kka}: a large $\xi$ naturally induces large higher dimensional operators that can in turn change the physical predictions. The coefficient $\alpha$  of a radiatively induced $\sqrt{-g}R^2$ term in the Lagrangian was shown in Ref.~\cite{Salvio:2015kka} to be subject to the following strong naturalness\footnote{Naturalness implies that condition because if one starts from an $|\alpha|$ much below that threshold the renormalization group flow generates an $|\alpha|$ comparable to it, unless very fine-tuned initial conditions are chosen.
% (see ... for the determination of the relevant $\beta$-functions)
 } bound   
\be |\alpha|\gtrsim \frac{\xi^2}{8\pi^2}.\label{natural-alpha0}\ee
%\xxx{Is the email of Odintsov relevant?}
A large value of $\xi$ is necessary at the classical level (see Eq.~(\ref{xi large}) and the corresponding discussion there). 
\noindent {\it However, in critical Higgs inflation $\xi$ does not have to be very large and a value of $\xi$ of order 10 is possible \cite{Hamada:2014iga,Bezrukov:2014bra,Hamada:2014wna}. Such value would lead to the trivial\footnote{This bound is trivial in view of the fact that an  $\alpha$ of order 1 leads to negligible corrections to Einstein gravity for energies much below the Planck scale (instead, for energies approaching this value we cannot have a model independent argument because Einstein gravity breaks down).} bound $\alpha \gtrsim 1$.} 
\noindent {\it Therefore, the critical Higgs inflation does not suffer from a fine-tuning of the high energy condition for the running couplings.}

 \begin{figure}[t]
\begin{center}
 \includegraphics[scale=0.55]{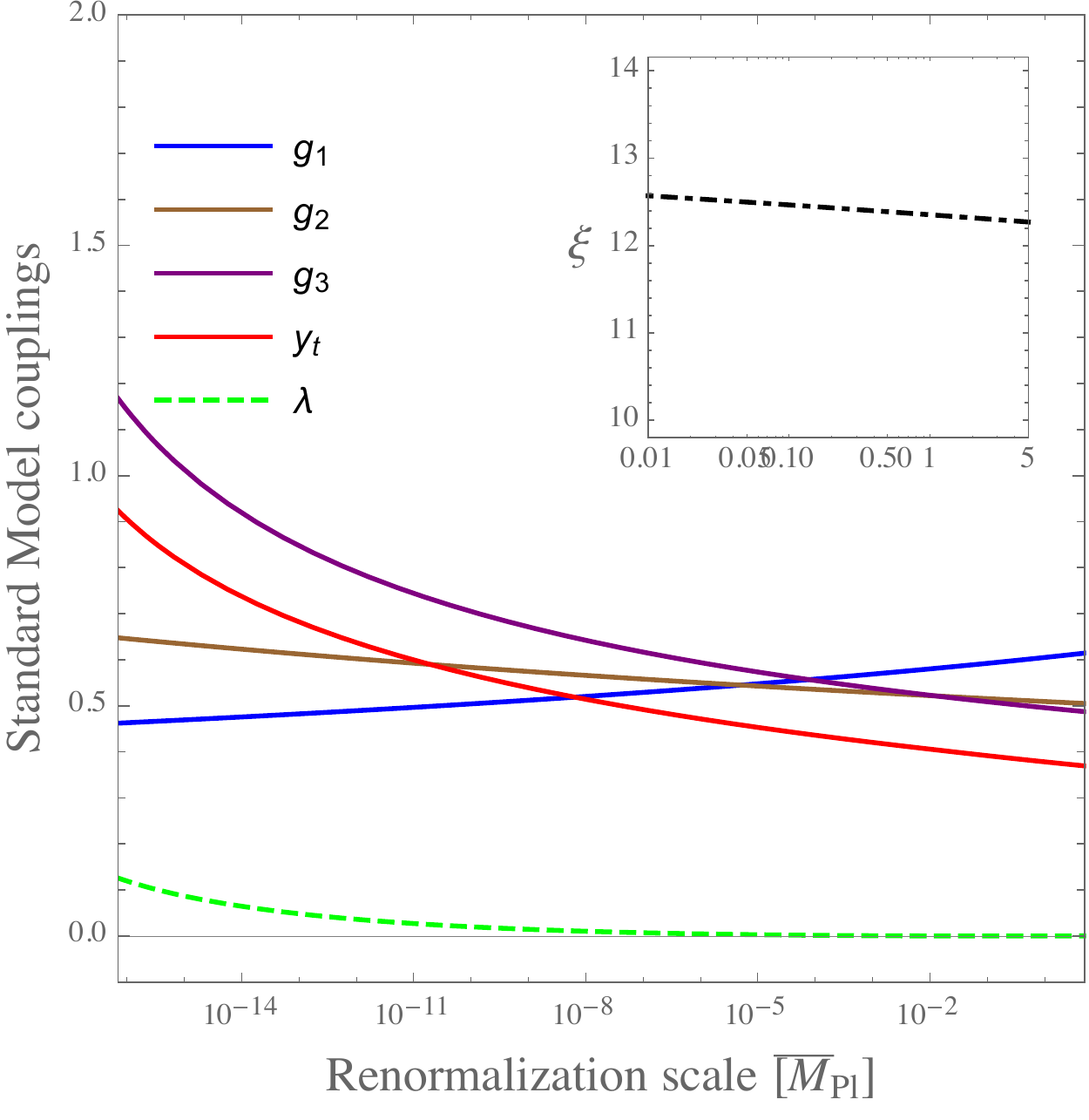}  \end{center}
 % \hspace{1cm}   \includegraphics[scale=0.6]{PDlambdaA0d05-y0d01-delta236.pdf}   
   \caption{\small \em Running of the SM couplings and $\xi$  for $M_t \approx 171.04$~GeV with the s-insertions.}
\label{running-potential1}
\end{figure}

\vspace{-0.2cm}

\section{Pre-inflationary dynamics}\label{Pre-inflationary dynamics}

Let us now analyze the dynamics of this system in the homogeneous case without making any assumption on the initial value of the time derivative $\dot \chi$. This analysis has been performed in \cite{Salvio:2015kka} for Higgs inflation at the classical level. Here we take into account quantum corrections to the potential. We will focus on the critical Higgs inflation for the reasons we discussed above.

In the Einstein frame $S_{\rm st}$  is 
\be S_{\rm st} =\int d^4x\sqrt{-g}\left[\frac{(\partial \chi)^2}{2}-U-\frac{\bp^2 }{2}R\right]   \ee
($U$ is the RG-improved Einstein frame effective potential).

 Let us assume a universe with 3-dimensional translation and rotation symmetries, that is a homogeneous and  isotropic FRW geometry. We do not regard this as a fine-tuning of the initial condition as it is implied by the requirement of having an enhanced symmetry (translation and rotation symmetries in this case). In other words, it is natural to assume that initially there was neither any special point in space nor any preferred direction. Furthermore, we will neglect  the spatial curvature in the FRW metric as the energy density is expected to be dominated by the inflaton during inflation.

 Then the equations for the spatially homogeneous field $\chi(t)$ and $a(t)$ are  
 \be \ddot\chi +\frac{\sqrt{3\dot\chi^2+6U}}{\sqrt2\bp}\dot\chi +U'  = 0\label{eq-k=0}\ee
 and 
\be \frac{\dot a^2}{a^2}=\frac{ \dot\chi^2+2U}{6 \bar M_{\rm Pl}^2}.  \label{EE1}\ee
From (\ref{eq-k=0}) and (\ref{EE1}) one can also derive the useful 
\be  \dot H= -\frac{\dot \chi^2}{ 2\bar M_{\rm Pl}^2}. \label{EE2}\ee
Note that, once we have a solution to Eq.~(\ref{eq-k=0}) we can immediately determine $a(t)$ through Eq.~(\ref{EE1}). Therefore, our job now is to solve  Eq.~(\ref{eq-k=0}) with appropriate initial conditions 
\be \Pi(\bar t) = \overline\Pi \qquad \chi(\bar t)=\overline \chi, \ee 
where $\bar t$ is some initial time before inflation and $\overline \chi$ and $\overline \Pi$ are the initial conditions for the dynamical variables\footnote{The initial condition for $a$ is not needed as the normalization of $a$ does not have a physical meaning for vanishing spatial curvature: indeed, given a solution $a$ of (\ref{EE1}), $c\, a$ is also a solution, for any constant $c$.} at $t=\bar t$.

Since we do not want to commit ourselves to any UV completion of Einstein gravity, we confine our attention to the regime where  quantum gravity  corrections are expected to be small,
\be  U \ll \bp^4 , \quad \dot\chi^2\ll \bp^4,   \label{smallQG}\ee
such that we can ignore the details of the UV completion. However, we do not require here to be initially in a slow-roll regime. The conditions in (\ref{smallQG}) come from the requirement that the energy-momentum tensor be small (compared to the Planck scale) so that a large curvature is not generated. The first condition in (\ref{smallQG}) is automatically fulfilled by the Higgs inflation potential: the quartic coupling $\lambda$ is small \cite{Bezrukov:2012sa, Buttazzo:2013uya,Andreassen:2017rzq,Salvio:2017eca}; note that $\lambda$  is particularly small in the critical HI \cite{Hamada:2014iga,Bezrukov:2014bra,Hamada:2014wna}, which is our main interest here.  The second condition in (\ref{smallQG}) is implied by the requirement of starting from an (approximately) de Sitter space, which is maximally symmetric; therefore we do not consider that as fine-tuning of the initial conditions. Indeed, in de Sitter we have to set $\dot  H=0$, which  leads to $\dot \chi =0$ given Eq.~(\ref{EE2}). 

However, note  that we cannot start from an exact de Sitter space given Eq.~(\ref{eq-k=0}):  the potential $U$ is indeed not exactly flat. Given that the extra symmetries  of de Sitter space (besides rotation and translation symmetries) are anyhow broken, there is no remaining symmetry that forces the field kinetic energy to be small compared to the potential energy or that forces the inertial term in the inflaton equation to be negligible with respect to the friction term. This motivates our study of initial conditions with generic kinetic energy.

\vspace{-0.2cm}

\subsection{Analytic discussion}

\begin{figure}[t]
\begin{center}
\includegraphics[scale=0.6]{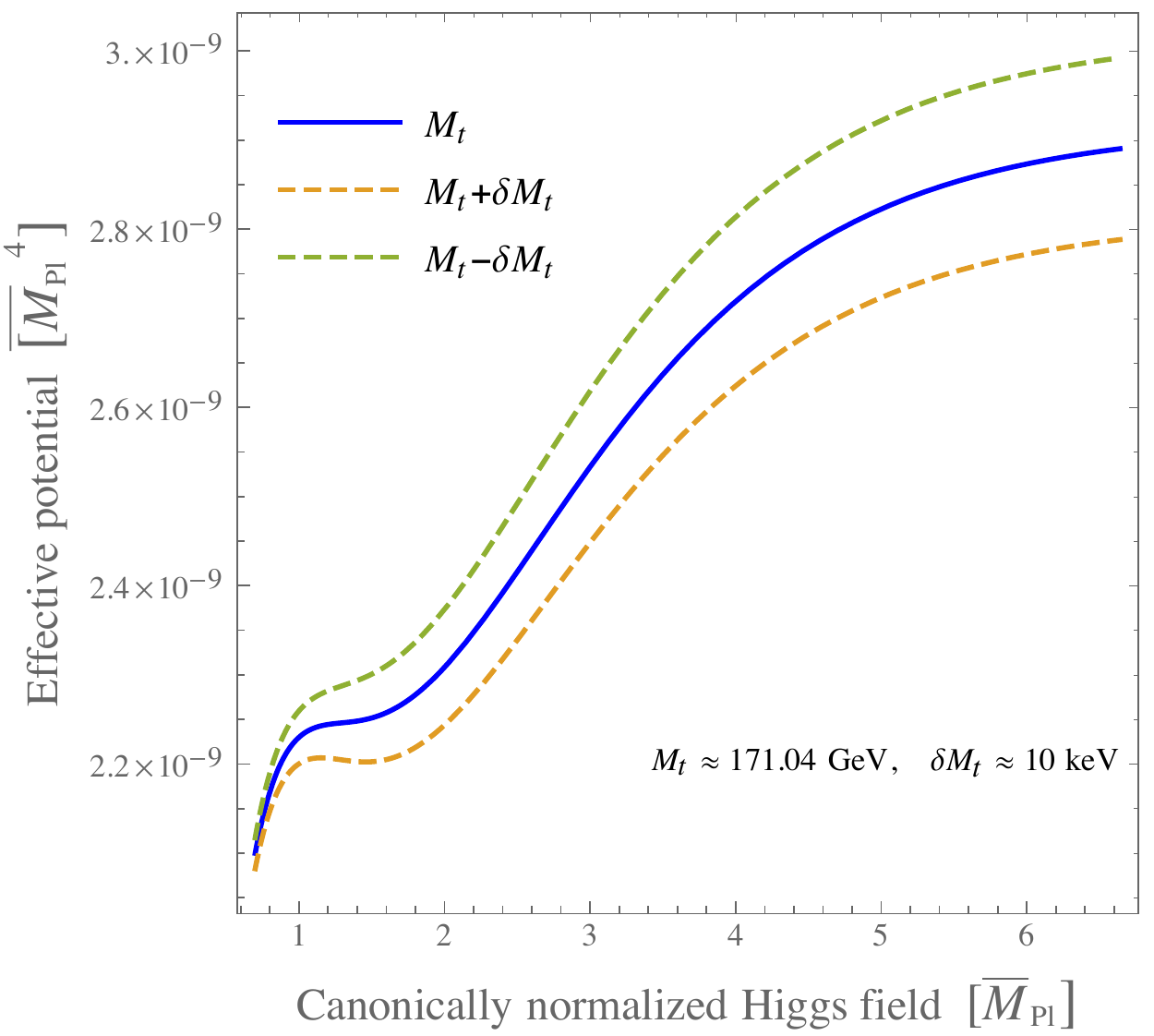}   
 \end{center}
 % \hspace{1cm}   \includegraphics[scale=0.6]{PDlambdaA0d05-y0d01-delta236.pdf}   
   \caption{\small \em SM effective potential (as defined in the text) with the $\xi$-coupling chosen as in Fig.~\ref{running-potential1}.}
\label{running-potential2}
\end{figure}
\begin{figure}[t]
\begin{center}
 \includegraphics[scale=0.55]{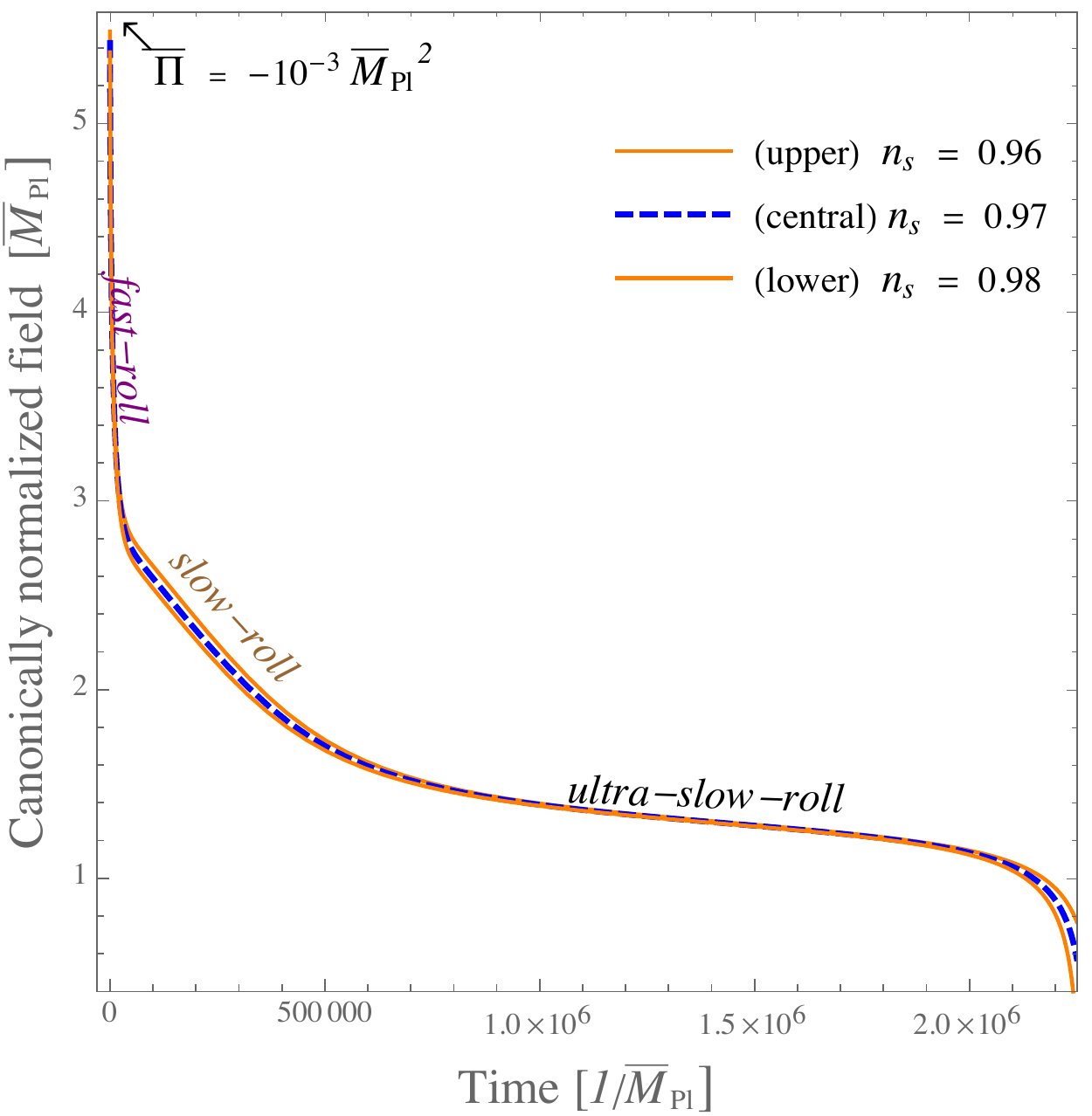}    
 \end{center}
 % \hspace{1cm}   \includegraphics[scale=0.6]{PDlambdaA0d05-y0d01-delta236.pdf}   
   \caption{\em \small Canonically normalized field $\chi$ as a function of time. The values of the parameters that are not quoted in the plot are chosen as in Figs. \ref{running-potential1} and \ref{running-potential2} and as explained in the text.}
\label{path}
\end{figure}
Since we do not know if and when exactly an inflationary slow-roll phase occurs it is essential for our purposes to have a description of the inflationary path that does not rely on the slow-roll approximation. Inflation in general takes place when 
\be \epsilon \equiv -\frac{\dot H}{H^2}  < 1, \ee 
which generalises  the usual definition of the slow-roll parameter $\epsilon_U$ (given in Eq.~(\ref{epsilon-def})) to situations where the slow-roll does not occur. In addition to $\epsilon$, one can introduce another parameter 
\be \delta \equiv  - \frac{\ddot\chi}{H \dot\chi}. \ee
If $\delta \ll 1$ one can neglect the inertial term in the inflaton equation (\ref{eq-k=0}) and reduce the problem to a single first order differential equation. The necessary and sufficient condition for inflation is only $\epsilon <1$. The condition $\delta \ll 1 $ is by no means necessary, although it leads to great simplifications. As we mentioned before and we will discuss in more detail in Sec.~\ref{Numerical studies}, in the critical Higgs inflation the condition $\delta \ll 1$ is not always satisfied because of the presence of an inflection point in the potential. Both because we will consider initial conditions with large kinetic energies and because of the inflection point, it is therefore important to solve the exact equation in  (\ref{eq-k=0}) without using the slow-roll approximation. Note that this also means that we cannot use now the slow-roll formula 
\be N=\int_{\phi_e}^{\phi_{\rm b}}\frac{U}{\bp^2}\left(\frac{dU}{d\phi}\right)^{-1}\left(\frac{d\chi}{d\phi}\right)^2d\phi \ee 
that we used before in Eq.~(\ref{e-folds}) to compute the 
number of e-folds $N$. We will use instead  the exact formula
\be N  = \int_{t_b}^{t_e} dt \,  H(t), \ee 
where $t_e$ is the time at the end of inflation and $t_b$ is the time when the inflationary observables $P_R$, $n_s$ and $r$ are measured.

One of the main purposes of the present work is to consider large initial kinetic energies, $\dot\chi^2 \gg U$, and study whether inflation is an attractor. This problem can be treated analytically during the first phase of the inflaton motion when $\dot\chi^2 \gg U$, such that the potential energy can be neglected. 
 In this case, combining Eqs.~(\ref{EE1}) and (\ref{EE2}) gives 
\be  \dot H+3H^2=0,\qquad (\dot\chi^2 \gg U),\ee
which  leads to 
\be  H(t)= \frac{\bar H}{1+3\bar H(t-\bar{t})},\qquad (\dot\chi^2 \gg U),\label{HlargeK}\ee
where $\bar H\equiv H(\bar t)$ and $\bar t$ is again some initial time. By inserting this result into Eq.~(\ref{EE2}) we find

\be \dot\chi^2= \frac{6\bp^2 {\bar H}^2}{\left[1+3\bar H(t-\bar{t})\right]^2},\qquad (\dot\chi^2 \gg U).\label{dotchiLargeK}\ee
This means that the kinetic energy density scales as $1/t^2$ if one takes into account the time dependence of $H$. This result \cite{Linde:1985ub} tells us that an initial condition with large kinetic energy is attracted towards one with smaller kinetic energy, but it also shows that neglecting the potential energy cannot be a good approximation for very large times.
 Moreover, notice that Eqs.~(\ref{HlargeK}) and (\ref{dotchiLargeK}) imply 
$ \ddot\chi = - 3H \dot\chi$,
so the dynamics is {\it not} approaching the usually assumed slow-roll condition $\delta \ll1$ that allows to drop the inertial term in the inflaton equation of motion.  Therefore, the argument above is not conclusive and  one needs to solve the equations taking into account $U$ in order to see if inflation is really an attractor.
We will do so numerically in the next subsection.

\subsection{Numerical studies}\label{Numerical studies}

In this section we present the numerical studies. We set here the reference value $\kappa \approx 2$
%2.02
for the parameter appearing in the optimal value of $\mub$ for the RG-improving, Eq.~(\ref{mubar}). As we will see, this leads to a realistic inflation.
%, given the level of approximation we are working on (see Sec.~\ref{Quantum corrections}).

% The dependence of the inflationary observables with respect to $\kappa$

In Fig.~\ref{running-potential1} we give the running of the largest SM couplings and $\xi$ with the s-insertions (see the discussion around Eq.~(\ref{s-form})). In that plot we use a value of $M_t$ close to criticality $M_t \approx 171.04$~GeV: the quartic coupling $\lambda$  nearly vanishes at high energies. There the minimum of $\lambda$ occurs at around $0.15 \bp$. In Fig.~\ref{running-potential2} we provide the corresponding  SM effective potential including the effect of $\xi$. We see that by varying  $M_t$ by only 10 keV around the critical value the potential changes significantly. This gives us an idea of the level of adjustment of $M_t$ required to have an inflection point in the potential, which is an important issue of the critical Higgs inflation. We could regard this either as a drawback or as an attractive feature of the model, depending on whether we regard this adjustment as a fine-tuning or a prediction of the model. A possible problem here is the tension between the critical $M_t$ and the measured value: $M_t = 172.51 \pm 0.50\,$GeV (ATLAS) and $M_t=172.44\pm 0.49\,$GeV (CMS)~\cite{ATLAStop2017}, which is, separately, at the 2-3$\, \sigma$ level\footnote{If we had  quantized the system in the Jordan frame before performing the conformal transformation (a definition of the theory known as prescription II) we would have found an even stronger tension~\cite{Bezrukov:2009db,Bezrukov:2009-2}. This is our main reason to choose prescription I. The extension of these calculations to prescription II is, therefore, beyond the scope of this article, but we expect that similar qualitative properties can be found with the alternative prescription.}. If future measurements and calculations will confirm this difference, new physics could be invoked to reconcile the two values of $M_t$, such as the well-motivated scenario of Ref.~\cite{Salvio:2015cja}.

\begin{figure}[t]
\begin{center}
  \includegraphics[scale=0.45]{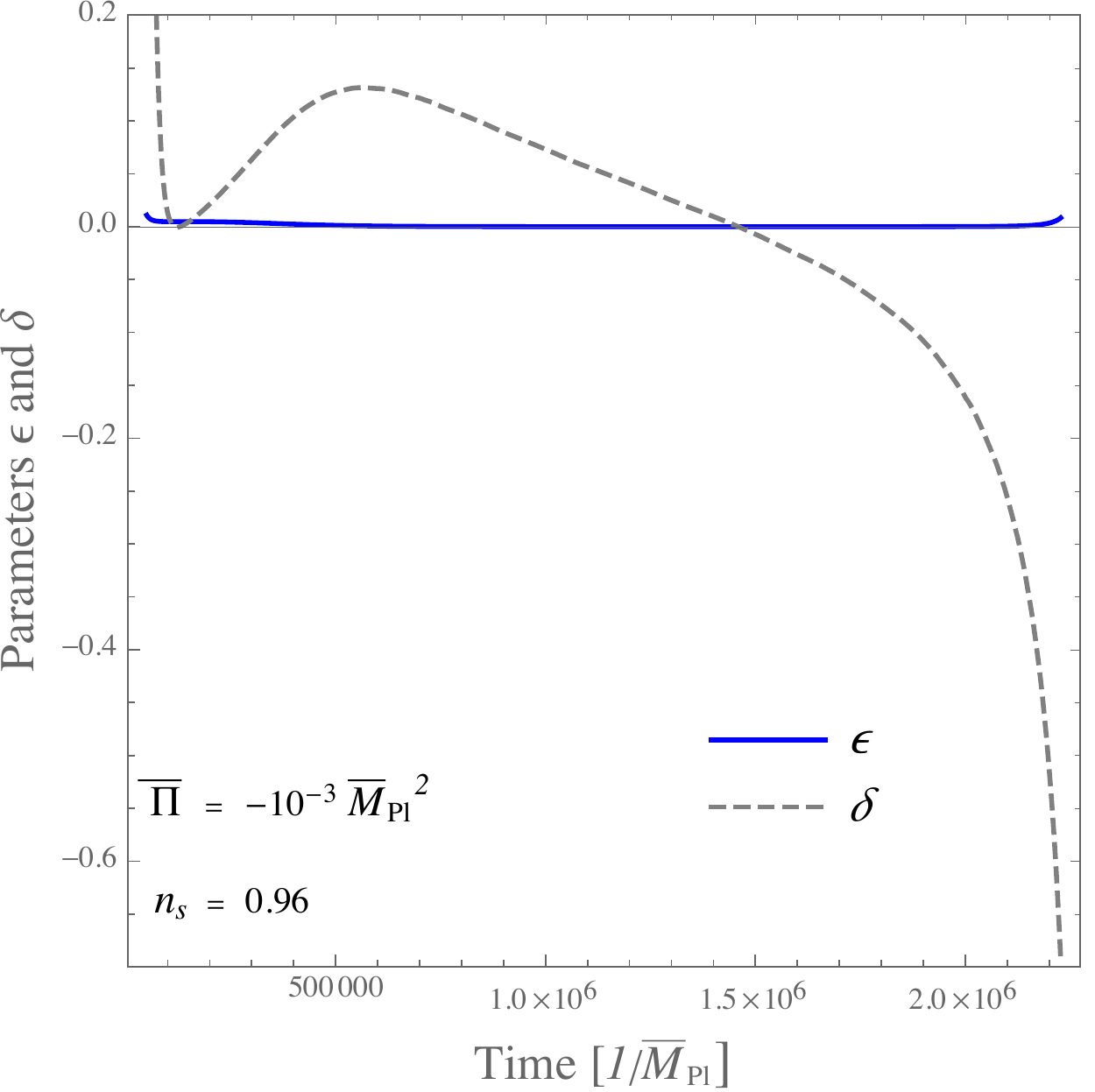}   
 \end{center}
 % \hspace{1cm}   \includegraphics[scale=0.6]{PDlambdaA0d05-y0d01-delta236.pdf}   
   \caption{\em \small The parameters $\epsilon$ and $\delta$ during a period of inflation. The values of the parameters that are not quoted in the plot are chosen as in Figs. \ref{running-potential1} and \ref{running-potential2} and as explained in the text.}
\label{epsilo-delta}
\end{figure}

We studied numerically the exact Higgs equation in (\ref{eq-k=0}). In Fig.~\ref{path} we provide the canonically normalized field $\chi$ as a function of time. We observe that even if start from a large kinetic energy\footnote{In that plot $\overline\Pi = - 10^{-3}\bp^2$, therefore the kinetic energy $\overline\Pi^2/2$ is much larger than the potential energy, as it can be checked by looking at Fig.~\ref{running-potential2}.} the field quickly reaches the slow-roll regime; the ultra-slow-roll regime~\cite{Kinney:2005vj,Germani:2017bcs} quoted in that plot corresponds to the period when the Higgs passed through the inflection point (see also Fig.~\ref{running-potential2}) where the potential is flatter.  In Fig.~\ref{epsilo-delta}  the parameters $\epsilon$ and $\delta$ are shown during a period of inflation. The main point of that plot is to show that, because of the inflection point~\cite{Ezquiaga:2017fvi} and the large initial kinetic energy, the parameter $\delta$ is not always very small, which indicates that  one could  not always neglect the inertial term in the inflaton equation during the whole inflation. The inflection point is reached at the time when $\delta =0$ as it can be checked by looking at Figs. \ref{running-potential2} and \ref{path}.

\begin{figure}[t]
\begin{center}
 \includegraphics[scale=0.47]{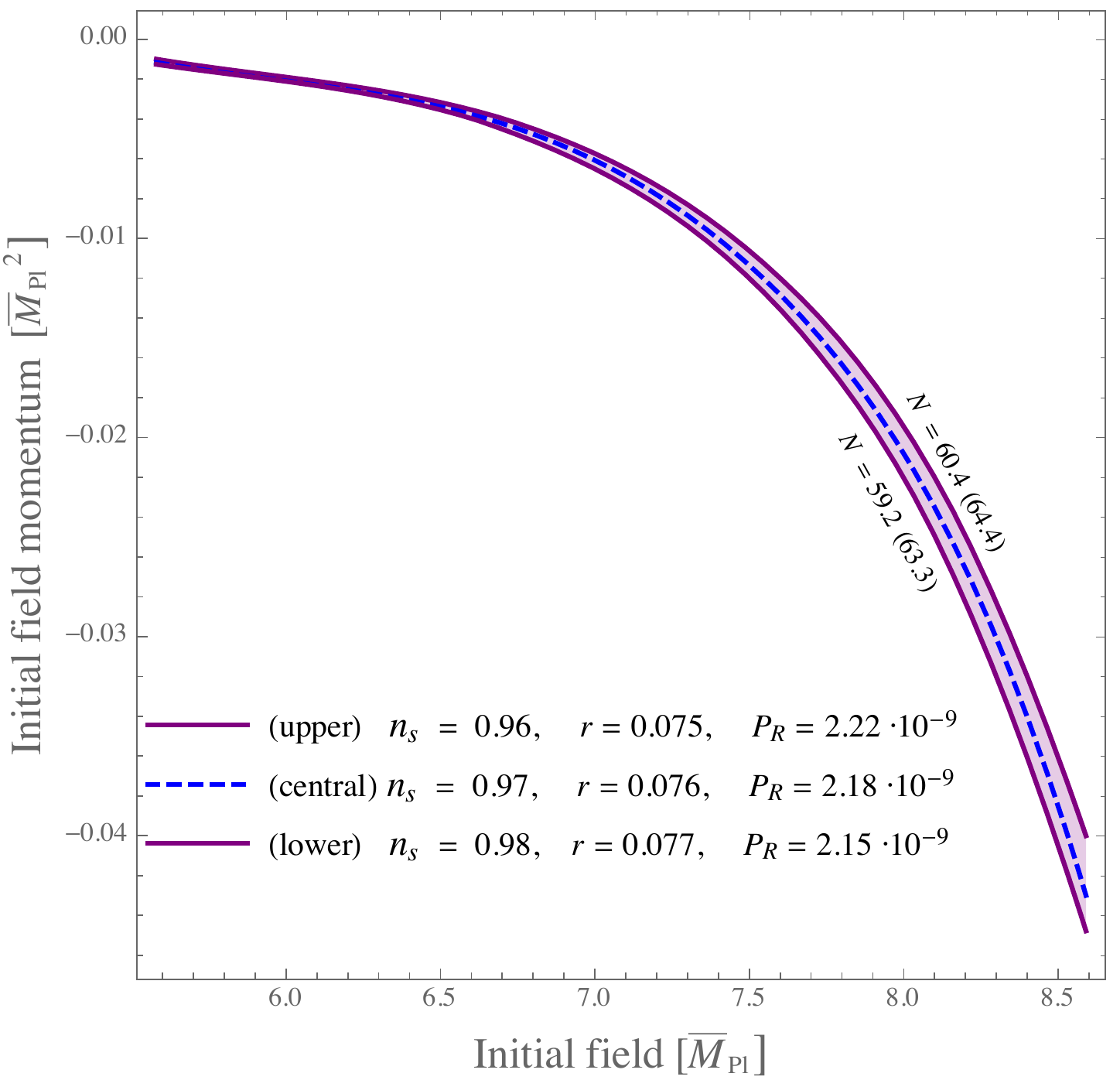}   
 \end{center}
 % \hspace{1cm}   \includegraphics[scale=0.6]{PDlambdaA0d05-y0d01-delta236.pdf}   
   \caption{\em   \small Initial conditions $\overline \chi$ and $\overline \Pi$ for the canonically normalized Higgs field $\chi$ and its momentum $\Pi\equiv\dot\chi$ respectively. The values of the parameters that are not quoted in the plot are chosen as in figs. \ref{running-potential1} and \ref{running-potential2} and as explained in the text. The values of inflationary parameters $N$ e-folds before the end of inflation are also provided (the values of $N$ inside the brackets indicate instead the total number of e-folds since the earliest time, when the initial conditions $\overline \chi$ and $\overline \Pi$ are given). 
   %The thickness of the lines corresponds to $2\sigma$ uncertainty in the value of the power spectrum, Eq.~(\ref{normalization}).
   }
\label{HIInitialCondition}
\end{figure}

Fig.~\ref{path} already indicates that inflation is an attractor in  the critical SM. We performed  a more general analysis by varying the initial momentum $\overline\Pi$ in Fig.~\ref{HIInitialCondition}. There, as well as in Figs.~\ref{path} and~\ref{epsilo-delta}, the initial  conditions for $\overline \Pi$  have been chosen to be negative because positive values favour inflation even respect to the case where the initial kinetic energy is much smaller than the potential energy:  this is because the potential, Eq.~(\ref{U}), is an increasing function of $\chi$ (for $\chi\gg v$).
We observe that a very large initial kinetic energy\footnote{In that plot we considered values of $\overline\Pi$ up to $-0.05\bp^2$ which corresponds to an initial kinetic energy density  of order $10^{-3}\bp^4$. We regard this value as the maximal kinetic energy density allowed to have negligible quantum gravity corrections (see also (\ref{smallQG})).} can be compensated by a very modest increase (not even of one order of magnitude) in the initial field value of the Higgs. This confirms that inflation is a strong attractor in this model. The situation is similar (and even slightly better) then the one of classical Higgs inflation \cite{Salvio:2015kka}  in this respect.  In the same plot we also show that the inflationary observables $n_s$, $r$ and $P_R$ are within the observational bounds~\cite{Ade:2015xua}.

Therefore, the critical Higgs inflation does not suffer from a fine-tuning problem for the initial conditions.

\vspace{-0.2cm}

\section{Conclusions}\label{Conclusions}

\vspace{-0.1cm}

In this paper we have studied whether Higgs inflation (HI) suffers from a fine-tuning of the high energy values of the parameters. In particular, it has been investigated the dependence of HI on the initial (pre-inflationary) conditions. In our analysis we assumed a spatially homogeneous and isotropic geometry pointing out the naturalness of this choice. As shown in \cite{Salvio:2015kka}, although the large-$\xi$ HI~\cite{Bezrukov:2007ep} does not suffer from any tuning of the initial conditions at the classical level, at the quantum level a  fine-tuning of the high energy values of some running parameters has to be performed, as discussed at the end of Sec.~\ref{Quantum corrections}. For this reason the main focus of this article has been critical HI~\cite{Hamada:2014iga,Bezrukov:2014bra,Hamada:2014wna}, which allows a drastic decrease of $\xi$. Moreover, critical HI, unlike the large-$\xi$ original version, has a single cut-off scale, $\bp$, where quantum gravity effects are expected to emerge, and is free from a much lower scale, where perturbative unitarity theory breaks down.

We pointed out that critical HI does not suffer from any fine-tuning of the high energy parameters, such as the one of large-$\xi$ HI noted in~\cite{Salvio:2015kka}. {\it The main result of this paper was that critical HI enjoys a robust inflationary attractor: even starting from a large kinetic energy density of the Higgs field (as compared to the potential energy density), the inflaton rapidly reaches the slow-roll behaviour.}

\vspace{-0.1cm}

\section*{Acknowledgments} 
\vspace{-0.1cm} 

\noindent   I thank J. Garcia-Bellido and M. Shaposhnikov  for useful discussions. This work was supported by the grant 669668 -- NEO-NAT -- ERC-AdG-2014.

\vspace{-0.cm} 

%\section*{References}

%\label{ciiiiiii}



%% References


%\bibliographystyle{elsarticle-num}


%\section*{References}
%\xxx{In the arXiv "References" appear twice}

\end{document}